\documentclass[journal]{IEEEtran}
\usepackage{amsfonts}
\usepackage{amssymb}
\usepackage{amsmath}
\usepackage{graphicx}
\usepackage{graphics}
\usepackage{color}
\usepackage{multirow}
\usepackage{cite}
\usepackage{bm}
\usepackage{algorithmicx}
\usepackage{algpseudocode}

\input{tcilatex}

\newtheorem{rema}{Remark}

\begin{document}

\title{Information-Coupled Turbo Codes for LTE Systems}
\author{
\IEEEauthorblockA {Lei Yang, Yixuan Xie, Xiaowei Wu, Jinhong
Yuan, Xingqing Cheng and Lei Wan} \vspace{-0.2cm} }
\maketitle

\begin{abstract}
We propose a new class of information-coupled (IC) Turbo codes to improve
the transport block (TB) error rate performance for long-term evolution (LTE) systems, while
keeping the hybrid automatic repeat request protocol and the Turbo decoder for each code block (CB)
unchanged. In the proposed codes, every two consecutive CBs in a
TB are coupled together by sharing a few common information bits. We propose
a feed-forward and feed-back decoding scheme and a windowed (WD) decoding
scheme for decoding the whole TB by exploiting the coupled information between CBs. Both decoding schemes
achieve a considerable signal-to-noise-ratio (SNR) gain compared to the LTE
Turbo codes. We construct the extrinsic
information transfer (EXIT) functions for the LTE Turbo codes and
our proposed IC Turbo codes from the EXIT functions of underlying
convolutional codes. An SNR gain upper bound of our proposed codes over the LTE
Turbo codes is derived and calculated by the constructed EXIT
charts. Numerical results show that the proposed codes achieve an SNR gain
of 0.25 dB to 0.72 dB for various code parameters at a TB error rate level
of $10^{-2}$, which complies with the derived SNR gain upper bound.
\end{abstract}


\section{Introduction}

In the long-term evolution (LTE) standards, \textit{transport block} (TB) based \textit{hybrid
automatic repeat request} (HARQ) is a key factor to provide low latency and
high speed data transmission \cite{LarmoComMag09}. In the TB based HARQ
protocol, a receiver uses only one bit acknowledgement (ACK) or negative
acknowledgement (NACK) to report the receiving status of a TB to the
transmitter. This mechanism minimizes the HARQ feedback overhead.
However, it results in a waste of transmission power and spectrum efficiency
since any code block (CB) errors in a TB will lead to the retransmission of the entire TB \cite{PaiTVT11}. In the current LTE standard, a TB can consist of tens of CBs \cite{LTEstd213}.
This number will further increase in the coming $5^{th}$ generation (5G) cellular networks
as the user peak throughput is expected to be increased by 100 -- 1000 times \cite{AndrewsJSAC14},
but the maximum CB length cannot be increased proportionally due to the high complexity for decoding long codes.
Therefore, the problem of wasting transmission power and spectrum efficiency associated with
the TB based HARQ protocol will become worse in the future.

To mitigate this problem, a \textit{CB based HARQ} scheme was proposed and
investigated in \cite{PaiTVT11} and \cite{LG3GPP166898} \cite{QC3GPP166375} for the 802.16e standard and the 5G new radio access technologies, respectively. In the scheme, only
erroneous CB (CBs) is (are) retransmitted. This saves transmission power and
improves spectrum efficiency, but leads to an excessive
amount of overhead in downlink/uplink control channels in order to manage a lot of
HARQ interlaces even within one TB \cite{QC3GPP166375}.

Another way to solve the aforementioned problem is to improve the TB
error rate (TBER) performance while keeping the HARQ protocol unchanged.
This can be straightforwardly achieved by using a much longer CB.
However, a much longer CB means a much longer decoding latency and
a much higher decoder complexity.

To exploit the benefits of long codes in terms of decoding threshold while
keeping the decoding latency and decoder complexity low, spatially-coupled
(SC) codes \cite{FelstromIT99} - \cite{LiangTCOM14} with windowed decoders
\cite{IyengarIT13} - \cite{IyengarIT11} are potential candidates.
Theoretically, codes with infinite length can be constructed by coupling
short codes in spatial domain. In practice, a low latency and low complexity
windowed decoder of finite length is employed to decode a terminated SC code
such that the decoding performance of the code approaches the theoretical
limit of the SC codes of infinite length \cite{IyengarIT13}.
Inspired by the promising performance of SC LDPC codes, authors in \cite%
{MoloudiISTC14} extended the spatial coupling technology to Turbo codes.
Belief-propagation (BP) decoding threshold analysis by density evolution
under binary erasure channel (BEC) has shown that BP thresholds of the SC Turbo
codes saturate to the maximum a posteriori (MAP) thresholds. Finite length
simulation results in \cite{AmatISWCS14} also showed that the SC Turbo codes
with windowed decoder have better decoding threshold than their non-SC
counterparts in BEC. Though the SC codes and windowed decoders are potential
solutions to improve the TBER performance for TB based HARQ schemes, the
window size of the windowed decoder has to be long enough to have a good
decoding threshold performance. Generally speaking, if the coupling memory
is $m$, the window size should be at least $m+1$. Therefore, the windowed
decoder has a much higher implementation complexity than the decoder for the
underlying non-SC counterpart.

It is understood that SC codes show a substantial coding gain compared to
their non-SC counterparts. Such a gain, namely convolutional gain, comes from
the fact that reliable messages at two ends of a terminated SC code are
continuously decoded and spread out, and gradually improve the quality of
other messages as iterative decoding progresses \cite{LentmaierIT10}.
Inspired by this reliable message spreading phenomenon, we propose to couple
the CBs in a TB into a chain and deliberately introduce a better decoding
threshold for the CBs at two ends of the chain. In \cite{WenGlobeCom2000},
the authors proposed a class of rate-compatible convolutional codes (CCs) by
inserting dummy bits into information sequence, which is termed
\textit{dummy bits inserting (DBI) CCs}. It has shown that the proposed
codes have comparable performance to the optimal repetition CCs with the
same code rate. In \cite{BreddermannTWC14}, the authors analyzed the DBI Turbo
codes by using \textit{extrinsic
information transfer} (EXIT) chart \cite{TenBrinkTcom01}. They showed that
the proposed Turbo codes outperform LTE Turbo codes in terms of frame error rate (FER) and
convergence speed.

Inspired by the reliable message spreading phenomenon of the SC codes \cite%
{LentmaierIT10} and the improved FER performance of the DBI Turbo codes over the LTE
Turbo codes in \cite{BreddermannTWC14}, we propose a new class of information-coupled (IC) Turbo
codes for LTE to improve the TBER performance. Meanwhile, we keep the TB
based HARQ protocol and the Turbo decoder for each CB in the LTE unchanged. The
main contributions are summarized below:

\begin{itemize}
\item We propose a new class of IC Turbo codes, which couple all CBs in a TB into a chain
by sharing a few common \textit{information bits} between every two consecutive CBs. Dummy bits
are inserted in the first and the last CBs of the coupled chain in order to
achieve a better decoding threshold and initial decoding performance of these two CBs than
that of other CBs. The advantages in these two
CBs can then be spread out to other CBs in iterative decoding
through the coupled information. The proposed IC Turbo codes are different from the SC
Turbo codes proposed in \cite{MoloudiISTC14} and \cite{AmatISWCS14}, where all CBs
are coupled together to form a much longer trellis. In our proposed IC Turbo
codes, each CB is terminated and the trellis is exactly the same as that of
the original LTE Turbo codes. Therefore, we can keep the Turbo decoder
for each CB unchanged. It is worth pointing out that we do not try to construct
a much longer code as that in the SC-LDPC codes.

\item We propose a feed-forward and feed-back (FF-FB) decoding scheme and a
windowed (WD) decoding scheme for our proposed IC Turbo codes to exploit
the coupled information between every two consecutive CBs. In the FF-FB decoding scheme,
the feed-forward (FF) decoding process traverses from the first CB in the
coupling chain to the last CB, and the feed-back (FB) decoding process conducts
in the opposite direction. These two decoding processes spread the reliable
messages from two ends of the coupled chain to other CBs and improve the
TBER performance. The WD decoding scheme goes through the coupled chain only
once from the first CB to the last CB, where each decoding window
consists of \textit{only two} consecutive CBs. The proposed IC
Turbo codes with both decoding schemes achieve a considerable TBER performance gain over the LTE
Turbo codes. The WD decoding scheme has a lower decoding
complexity than that of the FF-FB decoding scheme. Moreover, it keeps the on-the-fly
decoding feature, i.e., CBs can be decoded in a first-arrival-first-decode manner.

\item We propose a method to construct EXIT functions for the CC decoders of
the LTE Turbo codes with repetition from the EXIT functions of the underlying CC.
In the proposed method, the communication channel is modeled by two
independent parallel channels with different effective
signal-to-noise-ratios (SNRs). This is different from the method proposed in
\cite{BreddermannVTC11}, where each sub channel is modelled by an AWGN channel
cascaded with an erasure channel and the AWGN channel for all sub channels
has a same effective SNR. We derive an upper bound on the SNR gain of our IC Turbo codes over the LTE
Turbo codes and show that the proposed EXIT functions are
well fitted with simulation results. We also show that the derived upper
bound can effectively estimate the SNR gain of our proposed codes over the LTE
Turbo codes.

\item We evaluate the TBER performance of our proposed IC Turbo codes
through intensive Monte Carlo simulations. Simulation results demonstrate that our
proposed codes have a SNR gain ranges from 0.26 dB to 0.72 dB compared to the LTE
Turbo codes for various code parameters at a TBER level of $10^{-2}$.
\end{itemize}

\begin{figure*}[tbh]
\centering\includegraphics[width=5.5in]{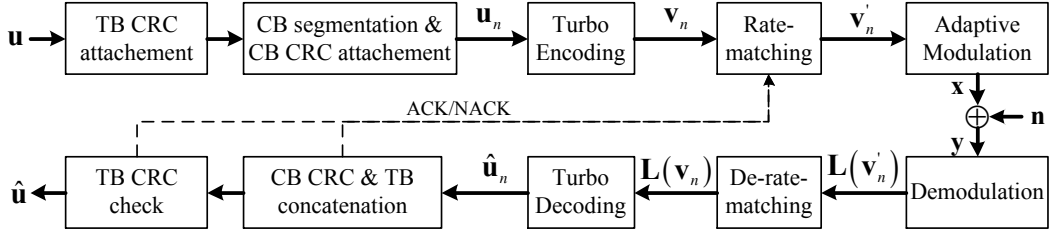} \centering
\caption{Block diagram of the HARQ process in LTE.}
\label{Fig_harq}
\end{figure*}

\section{TB Based HARQ in LTE and Problem Statement}

The TB based HARQ process in LTE is illustrated in Fig. \ref{Fig_harq}. A TB
$\mathbf{u}$ of length $L$ is fed to the physical layer for transmission.
The physical layer of a transmitter firstly attaches a 24-bits \textit{TB
cyclic redundancy check} (TB CRC) at the end of $\mathbf{u}$. If $L+24$ is
larger than the pre-defined maximum CB length, which is $6144$ in LTE
\cite{LTEstd212}, the TB is segmented into $N $ CBs and each CB is attached
with a \textit{CB CRC} of $24$ bits. Otherwise, segmentation and CB CRC
attachment are omitted and the TB consists of only one CB, i.e., $N=1$. The
resultant CBs $\left\{\mathbf{u}_n\right\}, n=1,\cdots,N$, are fed to a
systematic Turbo encoder of rate $R_0 = \frac{1}{3}$ sequentially. The
Turbo encoder consists of two $8$-state \textit{parallel concatenated CCs}
(PCCCs) with octal generator polynomials $\left(1,\frac{15}{13}\right)$ and
one internal interleaver. Denote the output codewords of the Turbo encoder
by $\left\{\mathbf{v}_n\right\}, n=1,\cdots,N$. The Turbo codewords $\mathbf{v}_n$ are sent to a
rate-matching device to obtain the required code rate. The resultant
codewords are denoted by $\mathbf{v}_n^\prime$. After adaptive modulation,
signals $\mathbf{x}$ are transmitted. The length of $%
\mathbf{x}$ is determined by the TB length $L$, the segmentation rules in
LTE and the modulation and coding scheme used for the TB. We omit these
details here because they are not essential to our proposed IC Turbo codes.
We refer the interested readers to \cite{LTEstd212}.

At the receiver side, noisy signals
\begin{equation}
\mathbf{y}=\mathbf{x}+\mathbf{n}  \label{Eq_rcv_signal}
\end{equation}%
are received, where $\mathbf{n}$ is an additive white Gaussian noise (AWGN)
vector with i.i.d components. Each component has zero mean and variance $%
\sigma_{ch}^2=\frac{N_0}{2}$, where $N_0$ is the double-sided noise power spectrum density.
Here, $\mathbf{y}$ and $\mathbf{n}$ have the same length as $\mathbf{x}$. SNR is defined as
$\rho=\frac{E_s}{N_0}$ and $E_s$ is the average symbol energy.

Upon receiving the noisy signals $\mathbf{y}$, a soft demodulator calculates
the log-likelihood ratio (LLR) for each coded bit by
\begin{equation}
L\left( v_{n,m}^{^{\prime }}\right) \triangleq \log _{2}\left( \frac{\Pr
\left( y_{m}\mid v_{n,m}^{^{\prime }}=0\right) }{\Pr \left( y_{m}\mid
v_{n,m}^{^{\prime }}=1\right) }\right) .  \label{Eq_LLR}
\end{equation}%
Here, $v_{n,m}^{^{\prime }}$ represents the $m$-th coded bit in the $n$-th codeword $%
\mathbf{v}^\prime_n$ and $y_{m}$ denotes the channel observation which contains $%
v_{n,m}^{^{\prime }}$. Then, $\mathbf{L}\left( \mathbf{v}_{n}^{^{\prime
}}\right) $ collects all $L\left( v_{n,m}^{^{\prime }}\right)$ and it is sent
to a de-rate-matching device to calculate the LLRs for coded bits in $%
\mathbf{v}_{n}$. In LTE, the rate-matching mechanisms include puncturing and
repetition. For the punctured bits in $\mathbf{v}_{n}$, we set the associated
LLRs to zeros. For a repeated coded bit $v_{n,m}$ in $\mathbf{v}_{n}$, if it
is repeated by $Q$ times and the associated LLRs for the $Q+1$ observations
are $\left\{ L\left( v_{n,m}^{^{\prime }}\right) ^{\left( 0\right) },\cdots
,L\left( v_{n,m}^{^{\prime }}\right) ^{\left( Q\right) }\right\} $, the LLR
for this coded bits is calculated by%
\begin{equation}
L\left( v_{n,m}\right) =\dsum\nolimits_{q=0}^{Q}L\left( v_{n,m}^{^{\prime
}}\right) ^{\left( q\right) }.
\end{equation}%
Here, we consider that the observations for the coded bit are independent.

After de-rate-matching, $\mathbf{L}\left( \mathbf{v}_{n}\right) $
collects all $L\left( v_{n,m}\right)$ for each CB, and then it is fed to a Turbo
decoder. The Turbo decoder consists of two constituent BCJR decoders \cite%
{BCJRIT79} and an interleaver/deinterleaver. By using an iterative decoding
process, estimated CBs $\left\{\hat{\mathbf{u}}_{n}\right\},n=1,\cdots ,N$,
are given. For each CB, if CB CRC detects an error, subsequent CBs in this
TB will not be processed and an NACK bit is sent by the receiver to its peer
transmitter. Otherwise, the estimated CBs $\left\{\hat{\mathbf{u}}%
_{n}\right\},n=1,\cdots ,N $, are concatenated and used to calculate the TB
CRC. If TB CRC detects an error, an NACK bit is sent by the receiver to its
peer transmitter, which triggers a retransmission process. Otherwise, an ACK
bit is sent and the TB is received successfully.

Note that, as only one bit feedback per TB is used in the LTE HARQ protocol,
the whole TB has to be retransmitted if any CB in the TB is in error. That
is why we call it \textit{TB based HARQ}. Obviously, when a TB consists of
several or tens of CBs, TB based feedback may result in a waste of
transmission power and a reduced transmission efficiency \cite{PaiTVT11} \cite{LG3GPP166898} \cite{QC3GPP166375}.%

We also note that each CB has a $24$-bits CRC if a TB consists of multiple
CBs. The CB CRC can be used as a CB-level early stopping criterion to reduce
decoding iterations of each CB. It can also be used as a TB-level early
stopping criterion to reduce the number of CBs to be decoded in an erroneous
TB. Both of them can save the receiver's computational resources \cite%
{ChengVTC08}. However, to enjoy the benefits of TB-level early stopping, CBs
must be decoded on-the-fly.

%\subsection{Problem Statement}

In this paper, we propose a new class of IC Turbo codes for LTE to improve
the TBER performance, i.e., decrease $\Pr \left( \hat{\mathbf{u}}\neq
\mathbf{u}\right) $. At the same time, we keep the TB based HARQ protocol
and the Turbo decoder for a CB in the current LTE standards unchanged. We mainly
consider the IC Turbo codes that have code rates $R_{IC}$ lower than the
mother code rate $R_0 = \frac{1}{3}$. In this case, repetition is
used as the rate-matching mechanism by the transmitter and chase-combining
is used by the receiver in LTE. Higher code rates can be realized by using the
same puncture mechanism as that used in the LTE Turbo codes.
In this paper, we mainly focus on the case $N\geq2$. For $N=1$, our codes
degrade to DBI Turbo codes as proposed in \cite{BreddermannTWC14}.

\section{Encoding of Our Proposed IC Turbo Codes}

In this section, we first present the encoding scheme of our proposed IC
Turbo codes. Then, the determination of code parameters for a given TB
length and required code rate is given. At last, the effective code rates
of the proposed codes are expressed w.r.t the mother code rate $R_{0}$.

\subsection{\label{Sec_encoding_scheme}Encoding Scheme of Our Proposed IC Turbo Codes}

\begin{figure}[tbh]
\centering\includegraphics[width=2.4in]{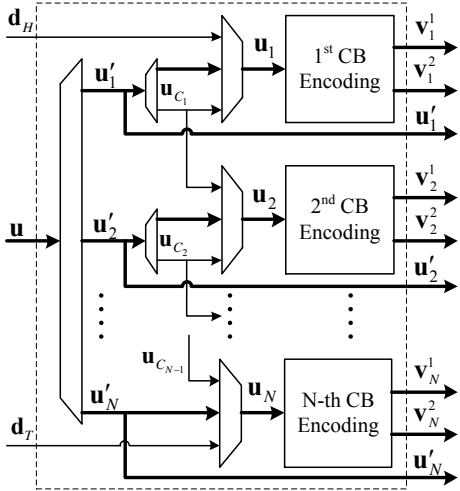} \centering
\caption{Block diagram of the encoding scheme for the proposed IC Turbo
codes.}
\label{Fig_encoding}
\end{figure}

The block diagram of the encoding scheme is shown in Fig. \ref{Fig_encoding}%
. In a big picture, the encoding process takes in a TB $\mathbf{u}$ of
length $L$ and two dummy bit sequences $\mathbf{d}_{H}$ and $\mathbf{d}%
_{T}$ of length $D_{H}$ and $D_{T}$ respectively, and generates $N$
systematic Turbo codewords $\left\{\mathbf{v}_{n}\right\}$, $n=1,\cdots ,N$.
Each codeword $\mathbf{v}_{n}$ consists of two parity check sequences $\mathbf{%
v}_{n}^{1}$ and $\mathbf{v}_{n}^{2}$, and a systematic bit sequence $\mathbf{%
u}_{n}^{\prime }$, where $\mathbf{v}_{n}^{1}$ and $\mathbf{v}_{n}^{2}$ are
generated by two constituent CC encoders in the Turbo code, respectively.

In the proposed encoder, several
common information bits are shared between two consecutive Turbo CBs, which can be exploited
in the iterative decoding to spread extrinsic information between CBs. We call this
type of codes \textit{information coupled Turbo codes}. This coupling method results
in a coupling memory $m=1$, which enables a low complexity windowed decoding scheme
in Section IV-B2. We can also increase the coupling memory, i.e, $m>1$,
however, it will result in a higher implementation complexity because a decoding
window of size $m+1$ is required to achieve a good decoding performance. In addition, dummy bits
are inserted in the first and the last Turbo CBs, which provide more reliable decoding outputs for
these two CBs than that of other CBs. These reliable messages
gradually spread out in the iterative decoding process and eventually
improve the overall TBER performance. The effect of these inserted dummy bits in
the FF-FB decoding scheme will be discussed in Section \ref{Sec_FF_FB}.

The detailed encoding process consists of three steps: 1) CB segmentation;
2) Information coupling and DBI; and 3) Turbo encoding, which are described
as below:

\textbf{\textit{Step 1} CB segmentation}: Segment TB $\mathbf{u}$ into $N$
\textit{information blocks}, i.e., $\mathbf{u}=\left[ \mathbf{u}_{1}^{\prime
},\cdots ,\mathbf{u}_{N}^{\prime }\right] $. Here, $\left[ \mathbf{u}^\prime%
_{1},\cdots ,\mathbf{u}^\prime_{N}\right] $ represents the concatenation operation
of $N$ vectors $\mathbf{u}^\prime_{1}$, $\mathbf{u}^\prime_{2}$ to $\mathbf{u}^\prime_{N}$. For $n = 1, \cdots, N-1$%
, let $D_{n}$ represent the number of shared information bits between CBs $%
\mathbf{u}_{n}$ and $\mathbf{u}_{n+1}$. Denote the length of the $n$-th CB $%
\mathbf{u}_{n}$ by $K_{n}$, $n\in\left\{1,\cdots,N\right\}$. Then, a TB is
segmented in such a way: let the first information block $\mathbf{u}%
_{1}^{\prime }$ have a length of $K_{1}-D_{H}$; let the $n$-th information
block $\mathbf{u}_{n}^{\prime }$, $n\in\left\{2,\cdots,N-1\right\}$, have a
length of $K_{n}-D_{n-1}$; and let the last information block $\mathbf{u}%
_{N}^{\prime }$ have a length of $K_{N}-D_{N-1}-D_{T}$.

\textbf{\textit{Step 2} Information coupling and DBI}: Construct $N$ CBs $%
\left\{\mathbf{u}_{n}\right\},n=1,\cdots ,N$, through information coupling
and DBI. Let $\mathbf{u}_{C_{n}}$ be the $D_{n}$ coupled information bits
between CBs $\mathbf{u}_{n}$ and $\mathbf{u}_{n+1}$, $n=1,\cdots ,N-1$,
i.e., the information bits in information block $\mathbf{u}_{n}^{\prime }$
shared by CBs $\mathbf{u}_{n}$ and $\mathbf{u}_{n+1}$. We call $D_n$ as
\textit{coupling length} between CBs $\mathbf{u}_{n}$ and $\mathbf{u}_{n+1}$%
. For the first CB $\mathbf{u}_{1}$, let $\mathbf{u}_{1}=\left\{ \mathbf{d}%
_{H},\mathbf{u}_{1}^{\prime }\right\} $, where $\mathbf{d}_{H}$ is the dummy bits
inserted into the first CB. For $n=1,\cdots, N-1$, let the $n$-th CB $\mathbf{u}%
%(head of the coupled chain)
_{n}=\left\{ \mathbf{u}_{C_{n-1}},\mathbf{u}_{n}^{\prime }\right\} $. For
the last CB $\mathbf{u}_{N}$, let $\mathbf{u}_{N}=\left\{ \mathbf{u}%
_{C_{N-1}},\mathbf{u}_{N}^{\prime },\mathbf{d}_{T}\right\} $, where $\mathbf{%
d}_{T}$ is the dummy bits inserted into the last CB.

\textbf{\textit{Step 3} Turbo Encoding}: A Turbo encoder, which consists of
two CC encoders and an interleaver, encodes the $N$ CBs. For a CB $\mathbf{u}%
_{n},n\in \left\{ 1,\cdots ,N\right\} $, the encoder generates the parity
check sequences $\mathbf{v}_{n}^{1}$ and $\mathbf{v}_{n}^{2}$ by two CC encoders
separately, and outputs $\mathbf{u}_{n}^{\prime }$, $\mathbf{v}_{n}^{1}$ and
$\mathbf{v}_{n}^{2}$ as a codeword. Note that, as the dummy bits $\mathbf{d}_{H} $ and
$\mathbf{d}_{T}$ are known by the receiver in advance, they do not need to be
transmitted through communication channels. As well as this, the coupled
information sequence $\mathbf{u}_{C_{n}},n\in \left\{ 1,\cdots ,N-1\right\} $,
is only transmitted in the $n$-th codeword $\mathbf{v}_{n}$ and are not
transmitted in the $\left( n+1\right) $-th codeword $\mathbf{v}_{n+1}$.

\begin{rema}
As shown in \cite{WenGlobeCom2000} that the distance spectrum of DBI CCs
depends on the number and the positions of these dummy bits in a CB. It is not
affected by the value of the dummy bits. Therefore, all-zero dummy bits are considered in
this paper, i.e., $\mathbf{d}_{H}=\mathbf{d}_{T}=\mathbf{0}$. In addition,
we consider that the dummy bits are equally spaced or nearly equally spaced
in a CB. The optimization of distributing the dummy bits in a CB is not considered
in this paper.
\end{rema}

\begin{rema}
The positions of coupled information bits in a CB will affect the decoding
performance of the proposed scheme. In this paper, we consider that the
coupled information bits are equally spaced or nearly equally spaced in the
associated CBs. The optimization of distributing the coupled information
bits in the associated CBs is not considered in this paper. We also consider
that the coupled information bits of a CB with its previous and next CBs are
independent, i.e., $\mathbf{u}_{C_{n-1}} \bigcap \mathbf{u}_{C_{n}}=\Phi$.
\end{rema}

\subsection{Code Parameters Determination and Effective Code Rate}

The encoding scheme described above introduced a set of code parameters $N$,
$K_{n}$, $D_{n}$, $D_{H}$ and $D_{T}$. Now we first present how to determine
these parameters for a given TB length $L$ and a given target effective
code rate $R_{IC}$. Then, we will discuss the relationship between the effective code rate of an IC
Turbo code to the mother code rate $R_0$.

Firstly, the segmentation process in Step 1 of the encoding scheme
guarantees
\begin{align}
L&=\left( K_{1}-D_{H}\right) +\dsum\nolimits_{n=2}^{N-1}\left(
K_{n}-D_{n-1}\right) \notag \\
&+\left( K_{N}-D_{N-1}-D_{T}\right) .
\label{Eq_TB_length}
\end{align}

Then, recall that the dummy bits $\mathbf{d}_{H}$ and $\mathbf{d}_{T}$ are
not transmitted in the output codewords and the coupled information sequences
$\left\{\mathbf{u}_{C_{n}}\right\},n=1,\cdots ,N-1$, are only transmitted once in Step 3 of the encoding scheme, the total output length of the proposed IC Turbo codes is written as%
\begin{align}
N_{IC}&=\left( \frac{K_{1}}{R_{0}}-D_{H}\right)
+\dsum\nolimits_{n=2}^{N-1}\left( \frac{K_{n}}{R_{0}}-D_{n-1}\right) \notag \\
&+ \left(
\frac{K_{N}}{R_{0}}-D_{N-1}-D_{T}\right) .  \label{Eq_TB_output_length}
\end{align}%
To satisfy the target effective code rate $R_{IC}$ for a given TB length $L$,%
\begin{align}
\frac{L}{R_{IC}}&=N_{IC}=\left( \frac{K_{1}}{R_{0}}-D_{H}\right)
+\dsum\nolimits_{n=2}^{N-1}\left( \frac{K_{n}}{R_{0}}-D_{n-1}\right) \notag \\
&+\left(\frac{K_{N}}{R_{0}}-D_{N-1}-D_{T}\right)  \label{Eq_TB_output_length_equal}
\end{align}%
should be satisfied.

Let $%
%TCIMACRO{\U{2124} }%
%BeginExpansion
\mathbb{Z}
%EndExpansion
^{+}$ be the set of positive integers and $\Omega $ be the set of valid CB
lengths in the LTE standard. Theoretically, any values of $N,$ $D_{n},$ $D_{H},$
$D_{T}\in
%TCIMACRO{\U{2124} }%
%BeginExpansion
\mathbb{Z}
%EndExpansion
^{+}$ and $K_{n}\in \Omega $ can be selected. In practice, we prefer to have
$K_{n}=K$ and maximize $K$ to exploit the best available coding gain. In
addition, we consider $D_{n}=D_{H}=D_{T}=D<\frac{K}{2}$ in this paper. Setting $%
D_{n}=D_{H}=D_{T}=D$ leads to a simple encoding structure and a similar CB
error rate (CBER) for all CBs in a TB, which will be shown in
Fig. \ref{Fig_FF-FB_saturation} in Section \ref%
{Sec_FF_FB}. Setting $D$ $<\frac{K}{2}$ guarantees the coupled information of a CB
with its previous and next CBs can be independent. By considering these
simplifications, (\ref{Eq_TB_length}) and (\ref{Eq_TB_output_length_equal})
become\footnote{%
In practice, one may not be able to select appropriate values for $N,$ $D\in
\mathbb{Z}^{+}$ and $K\in \Omega $ to satisfy (\ref{Eq_TB_length3}) and (\ref%
{Eq_TB_output_length_equal3}) for a given $L$ and $R_{IC}$. In that
case, zero padding to $\mathbf{u}$ is conducted to obtain a new TB of length $%
L^{^{\prime }}$ to guarantee that an appropriate set of $K,$ $N$ and $D$ can
be selected to satisfy (\ref{Eq_TB_length3}) and (\ref%
{Eq_TB_output_length_equal3}). In addition, $L^{^{\prime }}-L$ should be
minimized to save radio resources. Here, we assume (\ref{Eq_TB_length3}) and
(\ref{Eq_TB_output_length_equal3}) are satisfied for simplicity.}
\begin{align}
L=\left( N-1\right) \left( K-D\right) +\left( K-2D\right)
\label{Eq_TB_length3}
\end{align}%
and
\begin{align}
\frac{L}{R_{IC}}=N_{IC}=\left( N-1\right) \left( \frac{K}{R_{0}}%
-D\right) +\left( \frac{K}{R_{0}}-2D\right) .
\label{Eq_TB_output_length_equal3}
\end{align}%
Now, we propose to select the code parameters in such a way to maximize $K$ (equivalent to minimize $N$) and satisfy (\ref{Eq_TB_length3}) and (\ref{Eq_TB_output_length_equal3})
\begin{align}
N& =\min \left\{ N^{^{\prime }}:N^{^{\prime }}\in \mathbb{Z}^{+}\right\}
\notag \\
&\text{s.t. (\ref{Eq_TB_length3}) and (\ref{Eq_TB_output_length_equal3}) and
}D <\frac{K}{2},D\in\mathbb{Z}^{+},K\in \Omega ,  \notag \\
K& =\frac{L\left( 1-R_{IC}\right) R_{0}}{N\left( 1-R_{0}\right) R_{IC}},
\\
D& =\frac{NK-L}{N+1}.  \notag
\end{align}

For a given set of code parameters $N,$ $K$ and $D$, the effective code rate
of an IC Turbo code is calculated from (\ref{Eq_TB_length3}) and (\ref%
{Eq_TB_output_length_equal3}) as
\begin{align}
{R}_{IC}=& \frac{\left( N-1\right) \left( K-D\right) +\left( K-2D\right) }{%
\left( N-1\right) \left( \frac{K}{R_{0}}-D\right) +\left( \frac{K}{R_{0}}%
-2D\right) }  \notag \\
=& \frac{N\left( K-D\right) -D}{N\left( K-R_{0}D\right) -R_{0}D}R_{0}.
\label{Eq_code_rate}
\end{align}%
When the number of CBs $N\rightarrow \infty $, the effective code rate $%
R_{IC}\rightarrow \frac{K-D}{K-R_{0}D}R_{0}$; when the length of the coupled
information sequence $D\rightarrow 0$, $R_{IC}\rightarrow R_{0}$.

\section{Decoding of the Proposed IC Turbo Codes}

In the encoding scheme presented in Section \ref{Sec_encoding_scheme}, we
introduced coupled information between every two consecutive CBs. In this section, we
first present the decoding scheme for one CB, namely \textit{intra-CB
decoding}. Based on that, we propose two \textit{inter-CB decoding} schemes
to decode the whole TB by exploiting the coupled information between
every two consecutive CBs.

\subsection{\label{sec_intra_CB}Intra-CB Decoding}

\begin{figure}[tbh]
\centering\includegraphics[width=2.45in]{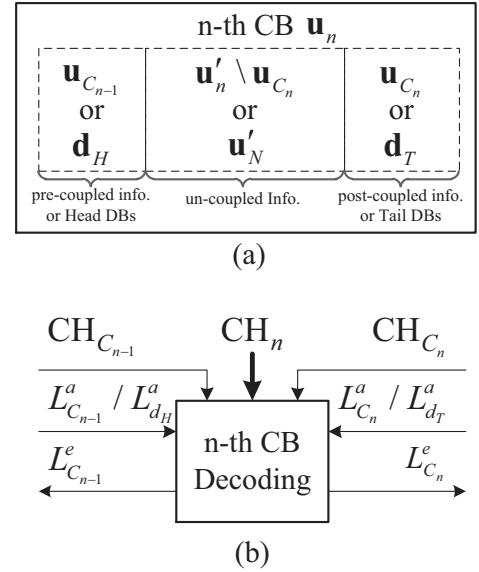} \centering
\caption{(a) Composition of a CB and (b) its associated decoding block
diagram.}
\label{Fig_decoding1}
\end{figure}

We can see from Fig. \ref{Fig_encoding} that each CB $\mathbf{u}_{n}$, $n\in
\left\{ 1,\cdots ,N\right\} $, consists of three parts of information, which
are illustrated in Fig. \ref{Fig_decoding1}(a). The first part is the
coupled information bits from the previous CB or the inserted dummy bits in the
first CB, i.e., $\mathbf{u}_{C_{n-1}}$ for CB $\mathbf{u}_{n}$, $n\in \left\{
2,\cdots ,N\right\} $, or $\mathbf{d}_{H}$ for CB $\mathbf{u}_{1}$. We call
this part of information as \textit{pre-coupled information} or \textit{head
dummy bits}, respectively. The second part is the information bits that are not
coupled with other CBs, i.e., $\mathbf{u}_{n}^{^{\prime }}\setminus \mathbf{u%
}_{C_{n}}$ for CBs $\left\{\mathbf{u}_{n}\right\}$, $n=1,\cdots ,N-1$, or $%
\mathbf{u}_{N}^{^{\prime }}$ for the last CB. Here $\mathbf{u}_{n}^{^{\prime }}\setminus \mathbf{u%
}_{C_{n}}$ represents the elements in $\mathbf{u}_{n}^{^{\prime }}$ exclude the
elements in $\mathbf{u}_{C_{n}}$. This part of information is termed \textit{%
un-coupled information}. The third part is the coupled information bits to
the next CB or the inserted dummy bits in the last CB, i.e., $\mathbf{u}_{C_{n}}$
for CB $\mathbf{u}_{n}$, $n\in \left\{ 1,\cdots ,N-1\right\} $, or $\mathbf{d}%
_{T}$ for CB $\mathbf{u}_{N}$. This part of information is named \textit{%
post-coupled information} or \textit{tail dummy bits}. As mentioned in Section
III-B that we consider $D<\frac{K}{2}$ in this paper, it guarantees each CB
is a composite of all three information parts.

According to the composition of CB $\mathbf{u}_{n}$, its associated decoding
block diagram is shown in Fig. \ref{Fig_decoding1}(b). Let \textbf{CH}$%
_{C_{n-1}}$ and $\mathbf{L}_{C_{n-1}}^{a}$ be the channel information and
the \textit{a priori LLR information} associated with the pre-coupled
information $\mathbf{u}_{C_{n-1}}$, \textbf{CH}$_{n}$ be the channel
information about the un-coupled information, and \textbf{CH}$_{C_{n}}$ and $%
\mathbf{L}_{C_{n}}^{a} $ be the channel information and the a priori LLR
information associated with the post-coupled information $\mathbf{u}_{C_{n}}$%
. Denote $\mathbf{L}_{C_{n-1}}^{e}$ and $\mathbf{L}_{C_{n}}^{e}$ the \textit{%
extrinsic LLR information} about the pre-coupled information $\mathbf{u}%
_{C_{n-1}}$ and the post-coupled information $\mathbf{u}_{C_{n}}$,
respectively. In addition, let $\mathbf{L}_{d_{H}}^{a}$ and $\mathbf{L}_{d_{T}}^{a}$
represent the a priori information of the head and the tail dummy bits. Recall that
we consider $\mathbf{d}_{H}=\mathbf{d}_{T}=\mathbf{0}$ in this paper, $%
\mathbf{L}_{d_{H}}^{a}=\mathbf{L}_{d_{T}}^{a}=\mathbf{\infty }$ since the
decoder has perfect knowledge about these dummy bits.

To decode CB $\mathbf{u}_{n}$, the Turbo decoder takes in channel
information CH$_{C_{n-1}}$, CH$_{n}$ and CH$_{C_{n}}$, and takes in the a
priori information $L_{C_{n-1}}^{a}$ ($L_{d_{H}}^{a}$) and $L_{C_{n}}^{a}$ ($%
L_{d_{T}}^{a}$) to perform intra-CB iterative decoding. When a predefined
intra-CB decoding stopping criterion is satisfied, such as a predetermined
maximum number of Turbo iterations has been reached or the CB CRC does not
detect an error, the decoder outputs $\widehat{\mathbf{u}}_{n}$ as an
estimation of CB $\mathbf{u}_{n}$. It also outputs the extrinsic information
$\mathbf{L}_{C_{n-1}}^{e}$ (except the first CB) and $\mathbf{L}_{C_{n}}^{e}$
(except the last CB), which will be used in the inter-CB decoding process between CBs.

Note that this intra-CB decoder is almost the same as that for the LTE Turbo
codes. The only difference is that the proposed decoder utilizes a priori
information from adjacent CBs and outputs extrinsic information to them.
This only affects the initialization and the outputs of the
decoder. Therefore, we can almost keep the LTE Turbo decoder for a CB
unchanged.

\subsection{Inter-CB Decoding}

Based on the intra-CB decoding scheme presented above, we propose two
inter-CB decoding schemes to exploit the extrinsic information between
coupled CBs to improve the TBER performance. The first decoding scheme,
namely \textit{FF-FB decoding} scheme, decodes serially from the first
undecoded CB to the last undecoded CB and then, if it is necessary, decodes
serially from the last undecoded CB to the first undecoded CB. This FF-FB
decoding process can be repeated for a few times to achieve an excellent
TBER performance. However, it suffers from a high decoding latency. As well
as this, for a long coupling length $D$, it has a high decoding complexity
(more details will be shown in Section V). To decrease the decoding latency
and the decoding complexity, we propose a \textit{WD decoding} scheme for
the proposed codes. It utilizes a window which consists of only two consecutive CBs and
slides from the first CB to the last CB. It retains good TBER performance
and has low decoding latency and decoding complexity. Next, we present the
proposed inter-CB decoding schemes.

\subsubsection{\label{Sec_FF_FB}FF-FB Decoding Scheme}

Let $\overrightarrow{\mathbf{L}_{C_{n}}^{e}}\left(\overrightarrow{\mathbf{L}%
_{C_{n}}^{a}}\right)$, $n\in\left\{1,\cdots,N-1\right\}$ denote the extrinsic (a priori)
information associated with $\mathbf{u}_{C_n}$, which is forwarded from the $n$%
-th CB to the $n+1$-th CB. Let $\overleftarrow{\mathbf{L}_{C_{n}}^{e}}$ $\left(
\overleftarrow{\mathbf{L}_{C_{n}}^{a}}\right)$ denote
the extrinsic (a priori) information sent from the $n+1$-th CB to the $n
$-th CB. The decoding scheme is described as below:

\textbf{\textit{Step 1} Initialize}: Let $\mathbf{L}_{d_{H}}^{a}=\mathbf{L}%
_{d_{T}}^{a}=\mathbf{\infty }$. Set the maximum intra-CB and inter-CB
decoding iterations to $I_{CB}$ and $I_{TB}$, respectively. Set the current
inter-CB iterations to $i_{TB}=0$.

\textbf{\textit{Step 2} FF Decoding}: The inter-CB decoder
decodes CBs serially from the first CB to the last CB. The forward
information $\overrightarrow{\mathbf{L}_{C_{n}}^{e}}$ $\left(\overrightarrow{\mathbf{L}_{C_{n}}^{a}}\right)
$, $n\in\left\{1,\cdots,N-1\right\}$, is calculated and passed down from
the $n$-th CB to the $n+1$-th CB. The details are as below:

If $i_{TB}=0$, for the $n$-th CB $\mathbf{u}_n$, the decoder uses the
associated channel information and the
a priori information $\overrightarrow{\mathbf{L}_{C_{n-1}}^{a}}$
about the pre-coupled information bits to estimate the CB. It also outputs the extrinsic information $\overrightarrow{\mathbf{L}_{C_{n}}^{e}}$ and $%
\overleftarrow{\mathbf{L}_{C_{n-1}}^{e}}$ associated with the post- and the pre-coupled information bits $\mathbf{u}_{C_{n}} $ and $\mathbf{u}_{C_{n-1}}$, respectively.

Note that, in this decoding iteration, the backward a priori information $%
\overleftarrow{\mathbf{L}_{C_{n}}^{a}}$ associated with the post-coupled information bits $\mathbf{u}_{C_{n}}$
is \textit{not available} for the $n$-th CB, i.e., $\overleftarrow{\mathbf{L}%
_{C_{n}}^{a}}=\mathbf{0}$. Therefore, only one round FF decoding cannot
fully exploit the coupled information between CBs.

If $i_{TB}\neq 0$, the inter-CB decoder decodes from the first undecoded CB to the
last undecoded CB by using
\textit{both} a priori information $\overrightarrow{\mathbf{L}_{C_{n-1}}^{a}}
$ and $\overleftarrow{\mathbf{L}_{C_{n}}^{a}}$ associated with the pre- and the post-coupled information bits to decode $\mathbf{u}_{n}$. Other
details are identical to that of $i_{TB}=0$.

Increase the current inter-CB iterations by one, e.g., $i_{TB}=i_{TB}+1$.

\textbf{\textit{Step 3} FF TB CRC}: Calculate TB CRC based on the
estimations $\left\{\widehat{\mathbf{u}}_{n}\right\},n=1,\cdots ,N$. If TB
CRC detects an error and $i_{TB}<I_{TB}$, go to Step 4; Otherwise, go to
Step 6.

\textbf{\textit{Step 4} FB Decoding}: The inter-CB decoder
decodes CBs serially from the last undecoded CB to the first undecoded CB.
The decoding of each CB is identical to the FF decoding with $i_{TB}\neq 0$.

Increase the current inter-CB iterations by one, e.g., $i_{TB}=i_{TB}+1$.

\textbf{\textit{Step 5} FB TB CRC}: Calculate TB CRC based on the
estimations $\left\{\widehat{\mathbf{u}}_{n}\right\},n=1,\cdots ,N$. If TB
CRC detects an error and $i_{TB}<I_{TB}$, go to Step 2; Otherwise, go to
Step 6.

\textbf{\textit{Step 6} Output decoded CBs}: Output $\left\{\widehat{\mathbf{%
u}}_{n}\right\},n=1,\cdots ,N$,  as the final estimations of $\left\{{\mathbf{%
u}}_{n}\right\}$.

\begin{rema}
The proposed FF-FB decoding scheme spreads reliable messages from the first
and the last CBs to other CBs via FF and FB decoding processes,
respectively. With a sufficiently large number of inter-CB decoding iterations, all CBs
gradually approach the decoding performance of the first and the last CBs,
which in turn improves the TBER performance. To show this reliable messages
spreading phenomenon, we plot the CBER against the CBs' indices
and the inter-CB iterations $i_{TB}$ in Fig. \ref{Fig_FF-FB_saturation}. The
TB consists of $17$ CBs, the coupling length $D=1024$, the maximum number of inter-CB
iterations is set to $I_{TB}=20$ and the channel SNR $\rho=-5.5$ dB.
Clearly, it can be seen that the first and the last CBs have a much better
CBER performance than that of other CBs in the first a few inter-CB
iterations. As the decoding process progresses, all CBs gradually approach a
similar CBER as that of the first and the last CBs.
\begin{figure}[h]
\centering\includegraphics[width=3.45in]{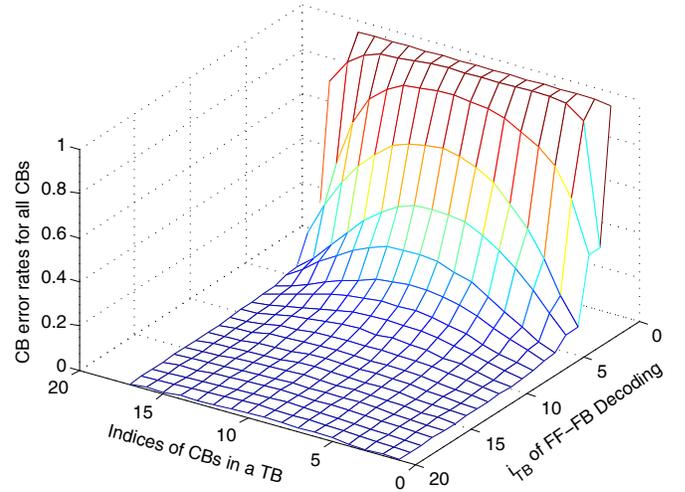} \centering
\caption{CBER performance of the proposed FF-FB inter-CB decoding
scheme against CBs' indices and the decoding iterations $i_{TB}$. The
parameters for the IC Turbo code are $K=6144$, $N=17$, $D=1024$, $%
L=14K$ and $R_{IC}=0.291$.}
\label{Fig_FF-FB_saturation}
\end{figure}

\end{rema}

The proposed FF-FB decoding scheme has a large TB decoding latency
as the decoding process has to go through all CBs for a few iterations to
achieve a good TBER performance. This also prevents using CB CRC as a TB-level
early stopping criterion to save the receiver's computational resources.
To decrease the decoding latency and the decoding complexity, we next propose a WD
decoding scheme.

\subsubsection{WD Decoding Scheme}
The proposed WD decoding scheme uses a window which consists of two consecutive
CBs to decode a CB and slides from the first CB to the last CB. The $n$-th decoding
window, which is used to decode the $n$-th CB, includes the $n$-th and the $\left(n+1\right)$-th
CBs. The decoding process inside the $n$-th window is described as below:

\textbf{\textit{Step 1} Initialize}: Obtain $\overrightarrow{\mathbf{L}%
_{C_{n-1}}^{a}}$, the a priori information associated with the pre-coupled
information of the $n$-th CB, from the $\left(n-1\right)$-th
decoding window. Let $\overleftarrow{\mathbf{L}_{C_{n}}^{e}}=\mathbf{0}$,
i.e., the $\left(n+1\right)$-th CB does not generate extrinsic information about
$\mathbf{u}_{C_n}$ yet. Let $\overleftarrow{%
\mathbf{L}_{C_{n+1}}^{a}}=\mathbf{0}$, i.e., without a priori information
from the $\left(n+2\right)$-th CB. Set the maximum iterations in the
decoding window to $I_{WD}$. Set the current decoding iterations $i_{WD}=0$.

\textbf{\textit{Step 2} Decode the $n$-th CB}: Let $\overleftarrow{\mathbf{L}%
_{C_{n}}^{a}}=\overleftarrow{\mathbf{L}_{C_{n}}^{e}}$, i.e., obtain the a priori
information associated with the post-coupled information $\mathbf{u}_{C_{n-1}}$
from the $\left(n+1\right)$-th CB. Decode the $n$-th CB using the a priori information $%
\overrightarrow{\mathbf{L}_{C_{n-1}}^{a}}$ and $\overleftarrow{\mathbf{L}%
_{C_{n}}^{a}}$, and the channel information $\mathbf{CH}_{C_{n-1}}$, $%
\mathbf{CH}_{n}$ and $\mathbf{CH}_{C_{n}}$. Output the estimated $n$-th CB $%
\widehat{\mathbf{u}}_{n}$ and the extrinsic information $\overrightarrow{%
\mathbf{L}_{C_{n}}^{e}}$ associated with the post-coupled information $\mathbf{u}%
_{C_{n}}$.

Increase the current decoding iterations by one, i.e., $i_{WD}=i_{WD}+1$.

\textbf{\textit{Step 3} CB CRC for the $n$-th CB}: Calculate CB CRC based on
$\widehat{\mathbf{u}}_{n}$. If CRC detects an error and $i_{WD}<I_{WD}$, go
to Step 4; Otherwise, go to Step 5.

\textbf{\textit{Step 4} Decode the $\left(n+1\right)$-th CB}: Let $\overrightarrow{%
\mathbf{L}_{C_{n}}^{a}}=\overrightarrow{\mathbf{L}_{C_{n}}^{e}}$, i.e., obtain the
a priori information about $\mathbf{u}_{C_n}$ from the $n$-th CB. Decode the
$\left(n+1\right)$-th CB using the a priori information $\overrightarrow{\mathbf{L}%
_{C_{n}}^{a}}$ and $\overleftarrow{\mathbf{L}_{C_{n+1}}^{a}}$, and the
channel information $\mathbf{CH}_{C_{n}}$, $\mathbf{CH}_{n+1}$ and $\mathbf{%
CH}_{C_{n+1}}$. Output extrinsic information $\overleftarrow{\mathbf{L}%
_{C_{n}}^{e}}$ and $\overrightarrow{\mathbf{L}_{C_{n+1}}^{e}}$. Go to Step 2.

\textbf{\textit{Step 5} Output the estimation of the $n$-th CB}: If CB CRC
for $n $-th CB does not detect an error, output $\widehat{\mathbf{u}}_{n}$
as the final estimation of the $n$-th CB. Let $\overrightarrow{\mathbf{L}%
_{C_{n}}^{a}}=\overrightarrow{\mathbf{L}_{C_{n}}^{e}}$. Move decoding window
one CB to the right, i.e., move to the $\left(n+1\right)$-th CB.

If CB CRC for the $n$-th CB detects an error, stop decoding. Other CBs
will not be decoded.

\begin{rema}
In the decoding process for the $n$-th CB, the a priori information $%
\overleftarrow{\mathbf{L}_{C_{n+1}}^{a}}$ is always zero since
the $\left(n+1\right)$-th CB does not have the a priori information about $\mathbf{u}%
_{C_{n+1}}$ from the $\left(n+2\right)$-th CB. Therefore, the extrinsic information $%
\overleftarrow{\mathbf{L}_{C_{n}}^{e}}$ generated by the $\left(n+1\right)$-th CB is
inferior to that generated by the FB decoding process in the FF-FB decoding
scheme. This in turn results in a decoding performance loss compared to the
FF-FB decoding scheme. This performance loss will be shown
in Section \ref{Sec_FF_FB_vs_WD}. However, the WD decoding scheme also achieves a
considerable SNR gain compared to the LTE Turbo codes. Moreover, it decodes CBs
on-the-fly, which leads to a low decoding latency and a low decoding complexity.
We will discuss the decoding complexity of the
proposed decoding schemes in the next section.
\end{rema}

\section{\label{Sec_complexity}Computational Complexity of the Proposed
Decoding Schemes}

In this section, we discuss the decoding complexity of the proposed decoding
schemes. As discussed in Section \ref{sec_intra_CB}, the intra-CB decoder
for our proposed scheme is almost the same as that for the LTE Turbo codes.
Therefore, we count the decoding complexity for a length-$K$ CB as an unit
and consider a length $L$ TB for the discussion of computational
complexity hereafter.

%\subsection{Maximum Computational Complexity for LTE Turbo Codes}

In the LTE standard, the TB is segmented into $\left\lceil \frac{L}{K}%
\right\rceil $ length-$K$ CBs\footnote{%
In the LTE standard, the TB segmentation rules result in $\left\lceil \frac{L%
}{K}\right\rceil $ CBs with similar lengths, but not necessarily exact the
same length of $K$. Here, we assume all CBs are of the same length for
simplicity.}, which results in a maximum computational complexity of $%
\left\lceil \frac{L}{K}\right\rceil $. Here, $\left\lceil x\right\rceil $
rounds up argument $x$ to the smallest positive integer. We use this maximum
computational complexity as a reference for that of our proposed decoding
schemes.

\subsection{Computational Complexity of the FF-FB Decoding Scheme}

In our proposed encoding scheme, the TB is segmented into $\left\lceil \frac{%
L+D}{K-D}\right\rceil $ CBs according to (\ref{Eq_TB_length3}).
Therefore, the maximum computational complexity for one FF-FB decoding
iteration is $\left\lceil \frac{L+D}{K-D}\right\rceil $. Consider a
maximum number of inter-CB iterations of $I_{TB}$, the overall computational
complexity is written as
\begin{equation}
\left\lceil \frac{L+D}{K-D}\right\rceil
\dsum\nolimits_{i_{TB}=0}^{I_{TB}-1}p_{i_{TB}}.  \label{Eq_complexity_FF_FB}
\end{equation}%
Here, $p_{i_{TB}}$ is the fraction of undecoded CBs in a TB in iteration $%
i_{TB}$. Obviously, $p_{0}=1$ because all CBs need to be decoded in the
first iteration.

\begin{figure}[h]
\centering\includegraphics[width=3.45in]{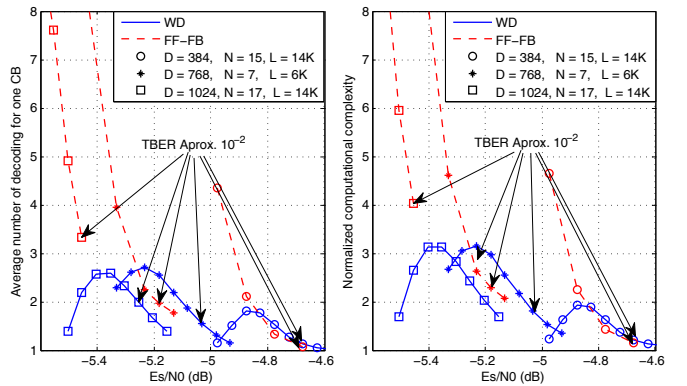} \centering
\caption{Average number of decoding for one CB (left hand side) and
normalized overall computational complexity (right hand side) for
coupling length $D=384, 768, 1024$ and TB length $L=14K, 6K, 14K$. $K=6144$.
}
\label{Fig_dec_complexity}
\end{figure}

We can see from (\ref{Eq_complexity_FF_FB}) that the overall computational
complexity of the proposed FF-FB decoding scheme is determined by $%
p_{i_{TB}},i_{TB}\in \left\{ 0,\cdots ,I_{TB}-1\right\} $, for a given $I_{TB}
$, which is in turn determined by the channel quality, the FER performance
of the underlying Turbo code and the number of CBs in a TB. Unfortunately, a
closed-form expression for the FER performance of the LTE Turbo code is not
available in general. As a result, we investigate the average number of
decoding for one CB and the normalized computational complexity by simulations
for various TB lengths $L$ and various coupling lengths $D$. Some results
are shown in Fig. \ref{Fig_dec_complexity}.

It can be seen from the left hand side of Fig. \ref{Fig_dec_complexity} that
for coupling length $D=384$ to $1024$, the average number of decoding for
one CB of the FF-FB decoding scheme grows from 1.05 to 3.3 at a TBER
level of $10^{-2}$. This is because when the coupling length $D$
increases, we need more inter-CB decoding iterations to fully exploit the
benefits provided by the coupled information. However, we will see in
Section \ref{sec_FF_FB_vs_bound} that the increased average number of decoding
iterations also increases the SNR gain of our proposed IC Turbo codes over the
LTE Turbo codes. The normalized overall computational complexity,
which is defined as the average number of decoding for one CB multiplied by $%
\left\lceil \frac{L+D}{K-D}\right\rceil /\left\lceil \frac{L}{K}%
\right\rceil $, is shown in the right hand side of Fig. \ref%
{Fig_dec_complexity}. The normalized overall
computational complexity ranges from 1.3 to 4 times of the maximum
computational complexity of the LTE Turbo codes when $D$ increases from $384$ to $1024$.

\subsection{Computational Complexity of the WD Decoding Scheme}

As in the FF-FB decoding scheme, the lack of closed-form FER expression of
the LTE Turbo code prevents us obtaining a closed-form expression of computational
complexity for the WD decoding scheme. Therefore, We also investigate its
computational complexity by simulations and show some results in Fig. \ref%
{Fig_dec_complexity}.

We can see from the left hand side of Fig. \ref{Fig_dec_complexity} that for
coupling length $D=384$ to $1024$, the average number of decoding for one CB
of the WD decoding scheme grows from 1.08 to 2 at a TB error rate level of $%
10^{-2}$. When $D=384$, the WD and FF-FB decoding schemes have a similar average
number of decoding per CB. However, when $D=768$ and $D=1024$, the WD decoding
scheme has a much lower average number of decoding iterations. The normalized
overall computational complexity for the WD decoding scheme is also shown in
the right hand side of Fig. \ref{Fig_dec_complexity}. Generally speaking,
for the coupling length $D=384$ to $1024$, the overall computational
complexity of the WD decoding scheme is about $1.35$ to $2.43$ times of that of the LTE Turbo codes.
Moreover, it can be seen that for various TB lengths and a relatively large coupling length, e.g., $D>384$, the WD decoding scheme has a much lower
computational complexity than that of the FF-FB decoding scheme. In particular, the WD decoding scheme saves $40\%$ computational resources compared to the FF-FB decoding scheme when $D=1024$.

\section{\label{Sec_EXIT}EXIT Chart Analysis}

In this section, we develop EXIT functions for the LTE Turbo codes
and our proposed IC Turbo codes. We also propose an upper bound for the SNR gain of our
proposed codes over the LTE Turbo codes, which can be calculated by
the developed EXIT functions.% In the following, capital letters and
%lower case letters are used to denote random variables and realizations of a
%random variable, respectively.

\subsection{Decoder Model and EXIT Chart Preliminaries}

EXIT chart is widely used to analyze the
convergence behavior of iterative decoders \cite{TenBrinkTcom01}. It evaluates the relationship
between the average extrinsic information $I_{E}$ generated by a
soft-input-soft-output (SISO) decoder and its input average a priori
information $I_{A}$. Plotting the EXIT functions of two constituent SISO decoders together, one
can clearly and intuitively predict the iterative decoding behavior of an
iterative decoder and observe the decoding trajectory.

\begin{figure}[h]
\centering\includegraphics[width=3.05in]{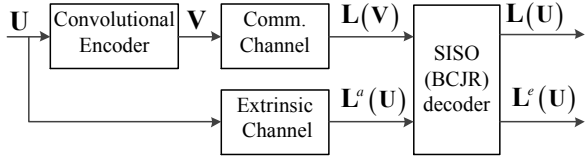} \centering
\caption{Decoding model for a component CC in Turbo codes.}
\label{Fig_decoder_model}
\end{figure}

To construct the EXIT function of a SISO decoder, a general information
theoretic decoder model was introduced in \cite{AshikhminIT04}. We adopt
this decoder model for a component CC in Turbo codes as shown in Fig. \ref%
{Fig_decoder_model}. Let $\mathbf{U}$ and $\mathbf{V}$ represent the random
variables with realizations of $\mathbf{u}_n$ and $\mathbf{v}_n$.
The SISO decoder, which is a BCJR decoder, uses channel observation LLRs $%
\mathbf{L}\left( \mathbf{V}\right) $ from \textit{communication channel} and
\textit{a priori} LLRs $\mathbf{L}^{a}\left( \mathbf{U}\right) $ delivered
by \textit{extrinsic channel} to compute the \textit{extrinsic} LLRs $%
\mathbf{L}^{e}\left( \mathbf{U}\right) $ as well as \textit{a posteriori}
LLRs $\mathbf{L}\left( \mathbf{U}\right) $. The LLRs are defined as in (\ref%
{Eq_LLR}). The communication channel models the real physical channel. It
also takes into account the rate-matching device and adaptive modulator at the
transmitter side, and the soft demodulator and de-rate-matching device at
the receiver side. The extrinsic channel is an artificial
channel which conveys the a priori LLRs $\mathbf{L}^{a}\left( \mathbf{U}%
\right) $ from other sources, such as another constituent decoder and the
decoders for the coupled CBs in our proposed IC Turbo codes.

\subsection{\label{Sec_EXIT_LTE}EXIT Functions for CCs in LTE Turbo Codes}

In LTE, code rates that lower than the mother code rate $R_{0} = \frac{1%
}{3}$ are achieved by repetition. Denote $R_{REP}<R_{0}$ the effective code
rate after repetition. According to the LTE rate-matching mechanism \cite{LTEstd212},
the maximum number of repetitions for a coded bit in $\mathbf{V}$ is written as%
% which is based on a circular buffer
\begin{align}
\Psi =\left\lceil \frac{R_{0}}{R_{REP}}\right\rceil -1.
\end{align}

Let $\mathbf{P}=\left[ p_{0},\cdots ,p_{\psi },\cdots ,p_{\Psi }\right] $
and $p_{\psi }$, $\psi \in \left\{ 0,\cdots ,\Psi \right\} $, be the fraction of
coded bits in $\mathbf{V}$\ that are repeated by $\psi $ times. Then
\begin{equation}
p_{\psi }=%
\begin{cases}
0, & \psi <\Psi -1 \\
1+\Psi -\frac{R_{0}}{R_{REP}}, & \psi =\Psi -1. \\
\frac{R_{0}}{R_{REP}}-\Psi, & \psi =\Psi%
\end{cases}
\label{Eq_rep_ratio}
\end{equation}

Eq. (\ref{Eq_rep_ratio}) means that there are at most two kinds of
bits in $\mathbf{V}$ which are repeated by $\Psi -1$ and $\Psi $ times.
Denote them by $\mathbf{V}_{\Psi -1}$ and $\mathbf{V}_{\Psi }$ respectively.
Consider a physical AWGN channel with SNR $\rho$, it is known that the effective SNR for a coded bit
is proportional to the number of repetitions for this bit, i.e., the
effective SNR is $\left( \psi +1\right) \times \rho$ for a coded bit
repeated by $\psi $ times. Assume a large enough channel interleaver,
$\mathbf{V}_{\Psi -1}$ and $\mathbf{V}_{\Psi }$ can be considered to
be transmitted through two independent communication channels with different
effective SNRs \cite{BreddermannVTC11} of $\rho_{\Psi -1}=\Psi \cdot \rho$ and
$\rho_{\Psi }=\left( \Psi +1\right) \cdot \rho$, respectively.

Now, let us consider the average information delivered by  the communication channel.
Let $\sigma _{ch}^{2}=\frac{E_{s}}{2\rho}$, $\sigma _{\Psi -1}^{2}=\frac{%
E_{s}}{2\rho_{\Psi -1}}=$ $\frac{E_{s}}{2\Psi \rho}$ and $\sigma _{\Psi
}^{2}$ $=\frac{E_{s}}{2\rho_{\Psi }}=$ $\frac{E_{s}}{2\left( \Psi +1\right)
\rho}$ be the noise variances of the physical AWGN channel, the
equivalent communication channel for $\mathbf{V}_{\Psi -1}$ and the
equivalent communication channel for $\mathbf{V}_{\Psi }$, respectively. We
convert these variances to the variances of their associated output LLRs
\cite{TenBrinkTcom01}, i.e., $\widetilde{\sigma }_{ch}^{2}=\frac{4}{\sigma
_{ch}^{2}}$, $\widetilde{\sigma }_{\Psi -1}^{2}=\frac{4}{\sigma _{\Psi
-1}^{2}}$ and $\widetilde{\sigma }_{\Psi }^{2}=\frac{4}{\sigma _{\Psi }^{2}}$%
. Then we have%
\begin{equation}
\widetilde{\sigma }_{\Psi -1}=\sqrt{\Psi }\widetilde{\sigma }_{ch}\text{ and
}\widetilde{\sigma }_{\Psi }=\sqrt{\Psi +1}\widetilde{\sigma }_{ch}.
\label{eq_equivalent_sigma}
\end{equation}

Let $I_{ch,\Psi -1}$ be the mutual information between $\mathbf{V}_{\Psi -1}$
and $\mathbf{L}\left( \mathbf{V}_{\Psi -1}\right) $, and $I_{ch,\Psi }$ be
the mutual information between $\mathbf{V}_{\Psi }$ and $\mathbf{L}\left(
\mathbf{V}_{\Psi }\right) $. The average information delivered by $\mathbf{L}%
\left( \mathbf{V}\right) $ about $\mathbf{V}$ is written as%
\begin{align}
I_{ch,REP} &=I_{ch,\Psi -1}p_{\Psi -1}+I_{ch,\Psi }p_{\Psi } \notag \\
&=J\left( \widetilde{\sigma }_{\Psi -1}\right) p_{\Psi -1}+J\left(
\widetilde{\sigma }_{\Psi }\right) p_{\Psi },  \label{Eq_apri_info_lte1}
\end{align}%
where \cite{TenBrinkTcom01}%
\begin{align}
J\left( \sigma \right) =1-\int_{-\infty }^{+\infty }\frac{e^{-\left(
\varepsilon -\frac{\sigma ^{2}}{2}\right) ^{2}/2\sigma ^{2}}}{\sqrt{2\pi }%
\sigma }\log _{2}\left( 1+e^{-\epsilon }\right) d\epsilon .
\end{align}%
Inserting (\ref{Eq_rep_ratio}) and (\ref{eq_equivalent_sigma}) into (\ref%
{Eq_apri_info_lte1}) results in%
\begin{align}
I_{ch,REP} & =J\left( \sqrt{\Psi }\widetilde{\sigma }_{ch}\right) \left( 1+\Psi
-\frac{R_{0}}{R_{REP}}\right) \notag \\
& + J\left( \sqrt{\Psi +1}\widetilde{\sigma }%
_{ch}\right) \left( \frac{R_{0}}{R_{REP}}-\Psi \right) .
\label{Eq_apri_info_lte2}
\end{align}

With the average channel information $I_{ch,REP}$, we now construct EXIT functions
for the LTE Turbo codes with repetition from that of the LTE mother Turbo code.
As the BCJR decoder is identical for both codes for the same CB length,
the decoder generates
identical extrinsic information if the channel information and the a priori
information received by the BCJR decoder are of the same value and have the
same distribution. Therefore, the EXIT functions for the LTE Turbo codes
with repetition can be derived by%
\begin{align}
\digamma _{REP}\left( I^{a}\left( \mathbf{U}\right) ,\widetilde{\sigma }%
_{ch}\right) =\digamma \left( I^{a}\left( \mathbf{U}\right) ,\widetilde{%
\sigma }_{ch}^{\prime }\right) .  \label{Eq_exit_lte_rep}
\end{align}%
Here, $\digamma _{REP}\left( I^{a}\left( \mathbf{U}\right) ,\widetilde{%
\sigma }_{ch}\right) $ is the EXIT function for the CC in an LTE
Turbo code with repetition, $\digamma \left( I^{a}\left( \mathbf{U}\right) ,\widetilde{%
\sigma }_{ch}^{\prime }\right) $ is the EXIT function for the CC in the LTE
Turbo mother code, $0\leq I^{a}\left( \mathbf{U}\right) \leq 1$ is the a
priori information conveyed by the extrinsic channel,
and $\widetilde{\sigma }_{ch}^{\prime }$ is calculated by%
\begin{align}
\widetilde{\sigma }_{ch}^{\prime } &=J^{-1}\left( I_{ch,REP}\right) \notag\\
 &=J^{-1}\left( J\left( \sqrt{\Psi }\widetilde{\sigma }_{ch}\right) p_{\Psi
-1}+J\left( \sqrt{\Psi +1}\widetilde{\sigma }_{ch}\right) p_{\Psi }\right) .
\label{Eq_equivalent_sigma_llr}
\end{align}

\begin{figure}[h]
\centering\includegraphics[width=3in]{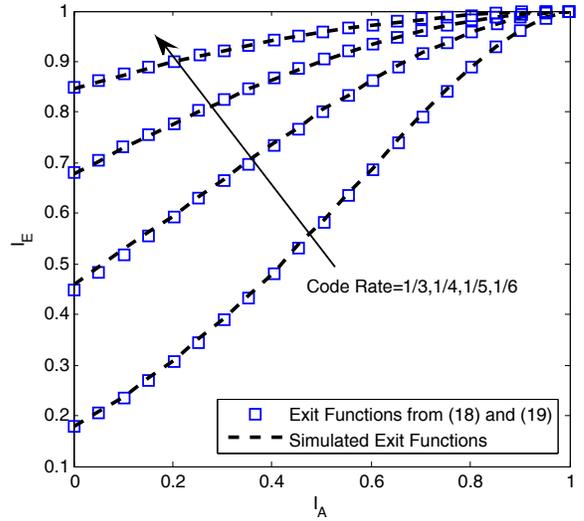} \centering
\caption{EXIT functions for CCs in LTE Turbo codes with repetition. The code
rates of CCs are $\frac{1}{3},\frac{1}{4},\frac{1}{5}$ and $\frac{1}{6}$.}
\label{Fig_Exit_LTE_REP}
\end{figure}

In (\ref{Eq_exit_lte_rep}) and (\ref{Eq_equivalent_sigma_llr}), we only
consider that the \textit{value} of channel information is identical for the LTE
Turbo codes with repetition and the LTE mother Turbo code. Strictly speaking, the
\textit{distribution} of the channel information should also be identical for the
BCJR decoder to generate the same output extrinsic information.
We generate the EXIT functions for the CCs in the LTE Turbo codes with repetition
by using (\ref{Eq_exit_lte_rep}) and (\ref%
{Eq_equivalent_sigma_llr}). We also obtain the EXIT functions through Monte
Carlo simulations based on random repetition\footnote{It has been shown in \cite{BreddermannVTC11} that the EXIT charts
obtained from random repetition can be used as a close approximation for the
deterministic repetition scheme used in LTE.}, i.e., $%
\mathbf{V}_{\Psi -1}$ and $\mathbf{V}_{\Psi }$ are selected randomly from $%
\mathbf{V}$. The results are shown in
Fig. \ref{Fig_Exit_LTE_REP}. It can be seen
that the EXIT functions constructed by our proposed method are well fitted
with the simulated results for various code rates.

\subsection{\label{Sec_EXIT_IC}EXIT Functions for CCs in our Proposed IC
Turbo Codes}

Recall that the intra-CB decoding scheme for the proposed IC
Turbo decoder takes in three parts of information to estimate a CB.
Therefore, the extrinsic channel for the CC decoder of the proposed IC Turbo codes consists of
three information channels: the extrinsic channel that conveys the a priori information $%
\mathbf{I}^{a,1}\left( \mathbf{U}\right) $ from the other constituent BCJR
decoder; the pre-coupled information channel that delivers the a priori
information $\mathbf{I}^{a,2}\left( \mathbf{U}\right) $ from the previous CB
(or dummy bits $\mathbf{d}_{H}$); and the post-coupled information channel that delivers
the a priori information $\mathbf{I}^{a,3}\left( \mathbf{U}\right) $ from
the post-coupled CB (or dummy bits $\mathbf{d}_{T}$). Note that, $\mathbf{I}^{a,2}\left(
\mathbf{U}\right) $ and $\mathbf{I}^{a,3}\left( \mathbf{U}\right) $ are
fixed in the intra-CB decoding process and evolve in the inter-CB decoding
process since we only exchange information between CBs in the inter-CB
decoding process.

As $\mathbf{I}^{a,2}\left( \mathbf{U}\right) $ and $\mathbf{I}^{a,3}\left(
\mathbf{U}\right) $ evolve with the inter-CB decoding progress, the EXIT
function for the CC decoder in our proposed IC Turbo code becomes a series
of EXIT functions. This results in a big challenge to calculate an exact
EXIT function for the CC decoder in our proposed scheme. Instead, we propose
to construct the EXIT function for the case where the pre-coupled
information bits and the post-coupled information bits are assumed to be
perfectly known by the decoder. Though this may over estimate the a priori
information $\mathbf{I}^{a,2}\left( \mathbf{U}\right) $ and $\mathbf{I}%
^{a,3}\left( \mathbf{U}\right) $, it simplifies the construction of EXIT
functions for our proposed codes. Moreover, it provides a decoding threshold
lower bound for our proposed IC Turbo codes. We propose to use this decoding
threshold lower bound to estimate the maximum SNR gain of our proposed
IC Turbo codes over the LTE Turbo codes. We call this maximum SNR gain
as \textit{SNR gain upper bound}. Simulation results show that the proposed SNR gain
upper bound has an accuracy within 0.1 dB for a set of code parameters, which
will be shown in Table \ref{Tbl_threshold} and Fig. \ref{Fig_FF-FB_vs_bounds}.

Now, we construct the EXIT\ function for the CC decoder in our proposed IC
Turbo codes by assuming perfect a priori information $\mathbf{I}^{a,2}\left(
\mathbf{U}\right) $ and $\mathbf{I}^{a,3}\left( \mathbf{U}\right) $. In \cite%
{BreddermannTWC14}, the authors proposed an effective way to construct the
EXIT chart for the CC decoder in DBI Turbo codes by shifting the EXIT chart
for the underlying CC decoder to an operation point with a higher a priori
information. It has been confirmed by simulation results that in the AWGN
channels, the EXIT functions constructed by the proposed method are well
fitted with the simulation results when the fraction of dummy bits
 is not higher than $40\%$. By assuming
perfect a priori information $\mathbf{I}^{a,2}\left( \mathbf{U}\right) $ and
$\mathbf{I}^{a,3}\left( \mathbf{U}\right) $, our proposed IC Turbo codes can
be viewed as a kind of DBI Turbo codes with fraction of dummy bits of $\frac{2D}{K}$. Thus, we adopt this method to
construct the EXIT function for the CC in our proposed IC Turbo codes.

As the pre-coupled and post-coupled information bits are assumed to be
perfectly known by the decoder, the a priori information $\mathbf{I}%
^{a,2}\left( \mathbf{U}\right) $ and $\mathbf{I}^{a,3}\left( \mathbf{U}%
\right) $ are written as%
\begin{align}
\mathbf{I}^{a,2}\left( \mathbf{U}\right) =\mathbf{I}^{a,3}\left( \mathbf{U}%
\right) =\frac{D}{K}.
\end{align}%
Since both $\mathbf{I}^{a,1}\left( \mathbf{U}\right) $ and $\mathbf{I}%
^{a,2}\left( \mathbf{U}\right) $ convey the information about the
pre-coupled information bits, the redundant a prior information $\frac{D}{K}%
\mathbf{I}^{a,1}\left( \mathbf{U}\right) $ should be removed from the total
a priori information $\mathbf{I}^{a}\left( \mathbf{U}\right)$. By the same
token, the redundant a prior information $\frac{D}{K}\mathbf{I}^{a,1}\left( \mathbf{%
U}\right) $ in both $\mathbf{I}^{a,1}\left(
\mathbf{U}\right) $ and $\mathbf{I}^{a,3}\left( \mathbf{U}\right) $ about
the post-coupled information bits should be removed. Recall that, we consider in this paper that
the pre-coupled information bits and the post-coupled information bits of a
CB are independent. Thus, the a priori information for the CC decoder is
written as%
\begin{align}
\mathbf{I}^{a}\left( \mathbf{U}\right) & =\mathbf{I}^{a,1}\left( \mathbf{U}%
\right) +\mathbf{I}^{a,2}\left( \mathbf{U}\right) +\mathbf{I}^{a,3}\left(
\mathbf{U}\right) \notag\\
&-\frac{D}{K}\mathbf{I}^{a,1}\left( \mathbf{U}\right) -%
\frac{D}{K}\mathbf{I}^{a,1}\left( \mathbf{U}\right)  \notag \\
& =\mathbf{I}^{a,1}\left( \mathbf{U}\right) +\frac{2D}{K}\left( 1-\mathbf{I}%
^{a,1}\left( \mathbf{U}\right) \right)  \label{Eq_apri_IC_Turbo}
\end{align}

Now, the EXIT function for the CC in our proposed IC Turbo codes can be
constructed by
\begin{align}
\digamma _{IC}\left( I^{a,1}\left( \mathbf{U}\right) ,\widetilde{\sigma }%
_{ch}\right) =\digamma \left( I^{a}\left( \mathbf{U}\right) ,\widetilde{%
\sigma }_{ch}\right) ,  \label{Eq_EXIT_IC_Turbo}
\end{align}%
where $\digamma _{IC}\left( I^{a,1}\left( \mathbf{U}\right) ,\widetilde{%
\sigma }_{ch}\right)$ is the EXIT function for the CC in an IC Turbo code and
$\digamma \left( I^{a}\left( \mathbf{U}\right) ,\widetilde{\sigma }%
_{ch}\right)$ is the EXIT function for the CC in the Turbo mother code. $%
0\leq I^{a,1}\left( \mathbf{U}\right) ,I^{a}\left( \mathbf{U}\right) \leq 1$
are related by (\ref{Eq_apri_IC_Turbo}) and $\widetilde{\sigma }^2_{ch} $ is
the variance of $\mathbf{L}\left( \mathbf{V}\right) $.

\begin{figure}[h]
\centering\includegraphics[width=3in]{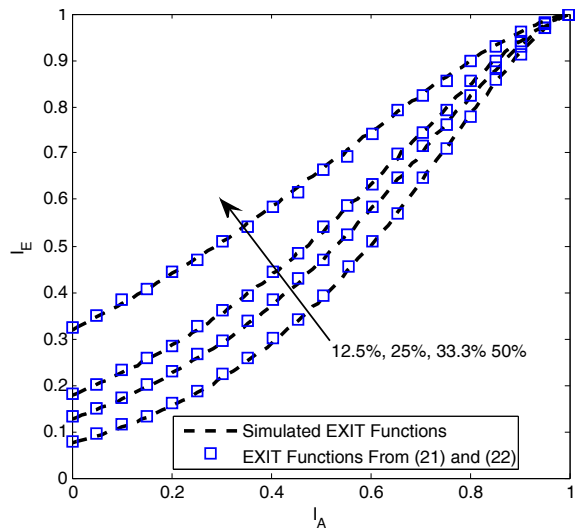} \centering
\caption{EXIT functions for CCs in our proposed IC Turbo codes with $12.5\%, 25\%, 33.3\%$ and $50\%$ coupled information bits.}
\label{Fig_Exit_IC_Turbo}
\end{figure}

We generate the EXIT functions for the CCs in our proposed IC Turbo codes
from that of the underlying CC by using (\ref{Eq_apri_IC_Turbo}) and (\ref%
{Eq_EXIT_IC_Turbo}). We also obtain the EXIT functions through Monte Carlo
simulations based on random DBI.
The EXIT functions for various coupling percentages are shown in Fig. %
\ref{Fig_Exit_IC_Turbo}. It can be seen that the EXIT functions constructed
by our proposed method agree with the simulation results well when the
percentages of coupled information bits are $12.5\%, 25\%, 33.3\%$ and $50\%$. %When $D=2048$, the constructed EXIT is not so accurate.

\section{\label{sec_numerical}Numerical Results}

In this section, we present the numerical results of our proposed IC Turbo
codes and the LTE Turbo codes under AWGN channels. We first compare the TBER performance of
the FF-FB and the WD decoding schemes for various IC Turbo codes.
Then the SNR gains of our proposed codes under the FF-FB decoding
scheme over the LTE Turbo codes for various TB lengths $L$ and coupling
lengths $D$ are investigated. We compare the simulated SNR gains with the
proposed SNR gain upper bounds to validate our proposed
EXIT chart functions. At last, we compare the TBER performance of our
proposed codes with the WD decoding scheme to that of the LTE Turbo codes for
various TB lengths $L$ and coupling lengths $D$.

In all simulations, the CB length $K=6144$ and the maximum number of intra-CB decoding
iterations $I_{CB}=8$ are considered. The maximum number of inter-CB iterations
is set to $I_{TB}=20$ for the FF-FB scheme and $I_{WD}=6$ for the WD decoding scheme, respectively.

\subsection{\label{Sec_FF_FB_vs_WD}Comparison of the FF-FB and WD Inter-CB
Decoding Schemes}

\begin{figure}[h]
\centering\includegraphics[width=3.25in]{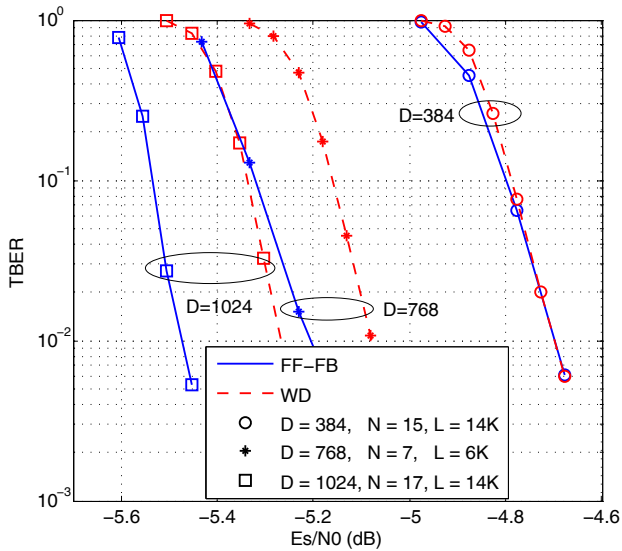} \centering
\caption{TBER performance of the proposed FF-FB and WD decoding schemes with
coupling length $D=384,768, 1024$ and TB length $L=14K, 6K, 14K$.}
\label{Fig_FF-FB_vs_WD}
\end{figure}

Fig. \ref{Fig_FF-FB_vs_WD} depicts the TBER performance of our
proposed FF-FB and WD inter-CB decoding schemes for three IC Turbo codes with $D=384, 768, 1024$ and $L=14K, 6K, 14K$, respectively. It can be seen from Fig. \ref{Fig_FF-FB_vs_WD} that for all codes, the FF-FB decoding scheme has a better TBER performance than that of the WD decoding scheme at the same SNR level. As discussed in Section IV-B Remark 4, this is because the
extrinsic information from the post-coupled CB for the WD decoding scheme is
inferior to that for the FF-FB decoding scheme. Moreover, it can be seen that the SNR gain of the FF-FB decoding scheme over the WD decoding scheme increases with the coupling length $D
$. This means the effect of lacking a priori information from the $\left(n+2\right)$-th CB
for the $n$-th decoding window becomes significant as the coupling length $D$ increases.

As the FF-FB decoding scheme has a better TBER at the same SNR level than that
of the WD decoding scheme, we will use the FF-FB decoding scheme to evaluate the
performance limit of our proposed codes next. On the other hand, the WD decoding scheme
has a lower decoding latency and a lower decoding complexity than that of the FF-FB
decoding scheme, we will use the WD
decoding scheme in Section \ref{sec_TBER_performance} to investigate the SNR
gains of our proposed codes over the LTE Turbo codes for practical purposes.

\subsection{\label{sec_FF_FB_vs_bound}Simulated SNR Gains vs the
Proposed SNR Gain Upper Bound}

We use the EXIT charts developed in Section \ref{Sec_EXIT} to calculate the
decoding thresholds of the LTE Turbo codes and our proposed codes for various
code rates. The proposed SNR gain upper bound in Section VI-C is calculated as the gap
between the decoding thresholds of a proposed code and a LTE Turbo code with
the same code rate. The calculated results are shown in Table \ref%
{Tbl_threshold}. It can be seen that the SNR gain upper bound increases with
the coupling length $D$.
\begin{table*}[tbp]
\caption{Decoding thresholds for the LTE Turbo codes and our proposed codes, and
the associated SNR gain upper bounds}
\label{Tbl_threshold}
\begin{center}
\begin{tabular}{|c|c|c|c|c|}
\hline
Code Rate & \multicolumn{2}{|c|}{Proposed Codes} & LTE Turbo Codes & SNR
Gain Upper Bound (dB) \\ \hline
& Coupling Length $D$ & Decoding Threshold (dB) & Decoding Threshold (dB) &
\\ \hline
0.318 & 384 & -5.22 & -4.92 & 0.3 \\ \hline
0.3 & 768 & -5.78 & -5.16 & 0.62 \\ \hline
0.286 & 1024 & -6.1 & -5.34 & 0.76 \\ \hline
\end{tabular}%
\end{center}
\end{table*}

\begin{figure}[h]
\centering\includegraphics[width=3.25in]{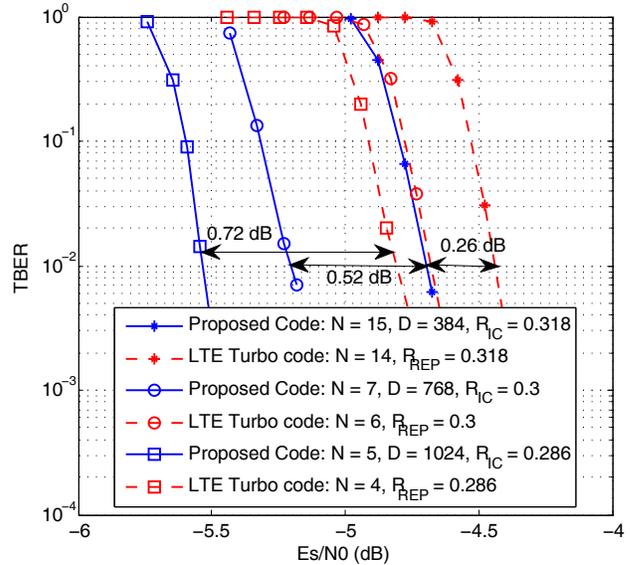} \centering
\caption{SNR gains of our proposed IC Turbo codes with coupling length $D=384, 768, 1024$ over
LTE Turbo codes for the same code rates.}
\label{Fig_FF-FB_vs_bounds}
\end{figure}

The simulated TBERs of the LTE Turbo codes and the
proposed codes with the FF-FB decoding scheme are demonstrated in Fig. \ref{Fig_FF-FB_vs_bounds}.
We can learn from the simulation results that for $D=384,768$
and $1024 $, the simulated SNR gains for various code rates are within 0.1 dB from the
proposed SNR gain upper bound at a TBER level of $10^{-2}$. This confirms that the
developed EXIT charts in Section \ref{Sec_EXIT} are valid and the proposed SNR gain upper bound is
tight for various coupling lengths. In addition, our simulation results
show that the proposed FF-FB decoding scheme for our proposed codes
can effectively exploit the benefits introduced by the coupled information
because it achieves the SNR gain upper bound within 0.1 dB.

\subsection{\label{sec_TBER_performance}TBER Performance of the Proposed IC
Turbo Codes under WD Decoding}

In this section, we evaluate the TBER performance of our proposed
codes with the WD decoding scheme and compare it to that of the LTE Turbo
codes with the same code rates. In particular, we
investigate the effect of the TB length $L$ and the coupling
length $D$ on the TBER performance of the proposed codes.

\begin{figure}[h]
\centering\includegraphics[width=3.25in]{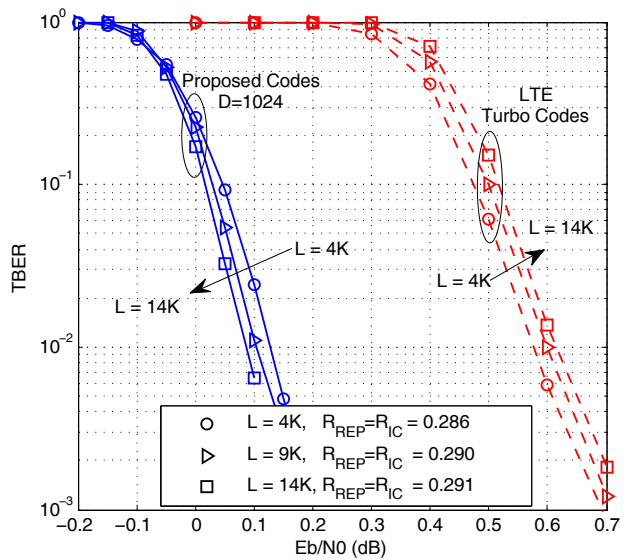} \centering
\caption{TBER of the proposed codes and the corresponding LTE Turbo codes
for TB length $L=4K, 9K, 11K$ and coupling length $D=1024$.}
\label{Fig_TBER_vs_N}
\end{figure}

Fig. \ref{Fig_TBER_vs_N} shows the TBER performance of the proposed codes
with $D=1024$ and $L=4K, 9K, 14K$, and that of the corresponding LTE turbo codes. We can see that our
proposed codes have considerable SNR gains over the LTE Turbo codes for various TB
lengths. When $L$ increases from $4K$ to $14K$, the SNR gain increases from 0.44 dB to 0.53 dB at
a TBER of $10^{-2}$ and increases from 0.43 dB to 0.5 dB at a TBER of $10^{-1}$.

\begin{figure}[h]
\centering\includegraphics[width=3.25in]{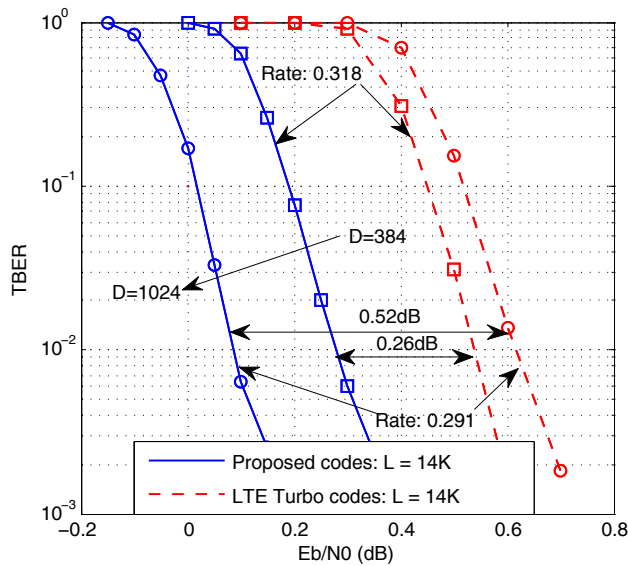} \centering
\caption{TBER of the proposed codes and the corresponding LTE Turbo codes
with coupling length $D=384, 1024$ and TB length $L=14K$.}
\label{Fig_TBER_vs_D}
\end{figure}

Fig. \ref{Fig_TBER_vs_D} shows the TBER performance of the proposed codes with
$L=14K$ and $D=384, 1024$, and that of the corresponding LTE turbo codes.
It can be seen from Fig. \ref%
{Fig_TBER_vs_D} that our proposed codes have significant SNR gains for
various coupling lengths $D$ compared to the LTE Turbo codes. Furthermore,
when coupling length $D$ increases from $384$ to $1024$
for the same TB length, the SNR gain increases from 0.26 dB to 0.52 dB
at a TBER of $10^{-2}$.

\section{Summary}

In this paper, we proposed a new class of IC Turbo codes by sharing
information bits between adjacent CBs. Two inter-CB decoding schemes are
proposed to exploit the coupled information introduced by the encoding
scheme. The proposed schemes achieve a considerable TBER performance improvement
compared to conventional LTE Turbo codes. New EXIT chart
construction method is proposed for the LTE Turbo codes from that of
underlying CC. An SNR gain upper bound of our proposed codes over the LTE
Turbo codes is derived by using our proposed EXIT
charts. Intensive simulation results show that the proposed codes have
considerable SNR gain compared to the conventional LTE Turbo codes.

%This paper introduced information coupling technology for TB based HARQ
%protocol to improve the TBER performance.
This information coupling technology can be extended to
other channel codes, such as LDPC codes. A few problems regarding
information coupling technology need to be addressed in the future. First of
all, the coupled information of a CB can be extended from adjacent CBs to
more CBs. In addition, the coupling length can be optimized
to further improve the TBER performance.

\end{document}